\documentstyle[aps,prl,preprint,floats,epsfig]{revtex}

\begin{document}

\preprint{\tighten\vbox{\hbox{\hfil CLNS 99/1625}
                        \hbox{\hfil CLEO 99-6}
}}

\title{Rare Decays of the $\eta '$}  

\author{CLEO Collaboration}
\date{\today}

\maketitle
\tighten

\begin{abstract}
We have searched for the rare decays $\eta ' \rightarrow e^{+}e^{-}\eta$
, $\eta ' \rightarrow e^{+}e^{-}\pi^{0}$, $\eta ' \rightarrow e^{+}e^{-}
\gamma$, and $\eta ' \rightarrow e\mu$ in hadronic events at the CLEO II 
detector.  The search is conducted on 4.80 fb$^{-1}$ of $e^{+}e^{-}$ 
collisions at 10.6 GeV center-of-mass energy at the Cornell 
Electron Storage Ring (CESR).  We find no signals in 
any of these modes, and set $90\%$ confidence level upper limits on their 
branching fractions: $B(\eta ' \rightarrow e^{+}e^{-}\eta) < 2.4\times 
10^{-3}$, $B(\eta ' \rightarrow e^{+}e^{-}\pi^{0}) < 1.4\times 
10^{-3}$, $B(\eta ' \rightarrow e^{+}e^{-}\gamma) < 0.9\times 10^{-3}$, and
$B(\eta ' \rightarrow e\mu) < 4.7\times 10^{-4}$.  We also investigated the
Dalitz plot of the
common decay mode $\eta ' \rightarrow \pi^{+}\pi^{-}\eta$.  We
fit the matrix element with the Particle Data Group parameterization
$\mid$M$\mid^{2} = $A$\mid 1 + \alpha y \mid ^{2}$, where $y$ is a linear 
function of the kinetic energy of the $\eta$, and find Re$(\alpha) = 
-0.021\pm 0.025$.
\end{abstract}
\newpage

{
\renewcommand{\thefootnote}{\fnsymbol{footnote}}

\begin{center}
R.~A.~Briere,$^{1}$
B.~H.~Behrens,$^{2}$ W.~T.~Ford,$^{2}$ A.~Gritsan,$^{2}$
H.~Krieg,$^{2}$ J.~Roy,$^{2}$ J.~G.~Smith,$^{2}$
J.~P.~Alexander,$^{3}$ R.~Baker,$^{3}$ C.~Bebek,$^{3}$
% ...
\end{center}
 
\small
\begin{center}
$^{1}${Carnegie Mellon University, Pittsburgh, Pennsylvania 15213}\\
$^{2}${University of Colorado, Boulder, Colorado 80309-0390}\\
$^{3}${Cornell University, Ithaca, New York 14853}\\
% ...
\end{center}

\setcounter{footnote}{0}
}
\newpage

The $\eta$ and $\eta '$ mesons share the same quantum numbers, and can both
be used to investigate C and CP violation, leptoquarks, lepton family 
violation, chiral perturbation theory, and other topics \cite{nefkens}.  But
while the $\eta$ has been the subject of several experiments, our
experimental knowledge of the $\eta '$ is limited;
many measured upper limits for its
decays are at the percent level \cite{pdg}.  In this Letter we present
searches for the rare decays $\eta ' \rightarrow e^{+}e^{-}\eta$, $\eta '
\rightarrow e^{+}e^{-}\pi^{0}$, $\eta ' \rightarrow e^{+}e^{-}\gamma$, and
$\eta ' \rightarrow e\mu$ produced in $e^{+}e^{-}$ collisions 
at 10.6 GeV center-of-mass energy.  We have concentrated on decays involving 
at least one electron because combinatoric backgrounds in hadronic events are
much larger for photons and charged pions than for leptons.

\begin{figure}
\centerline{\epsfig{figure=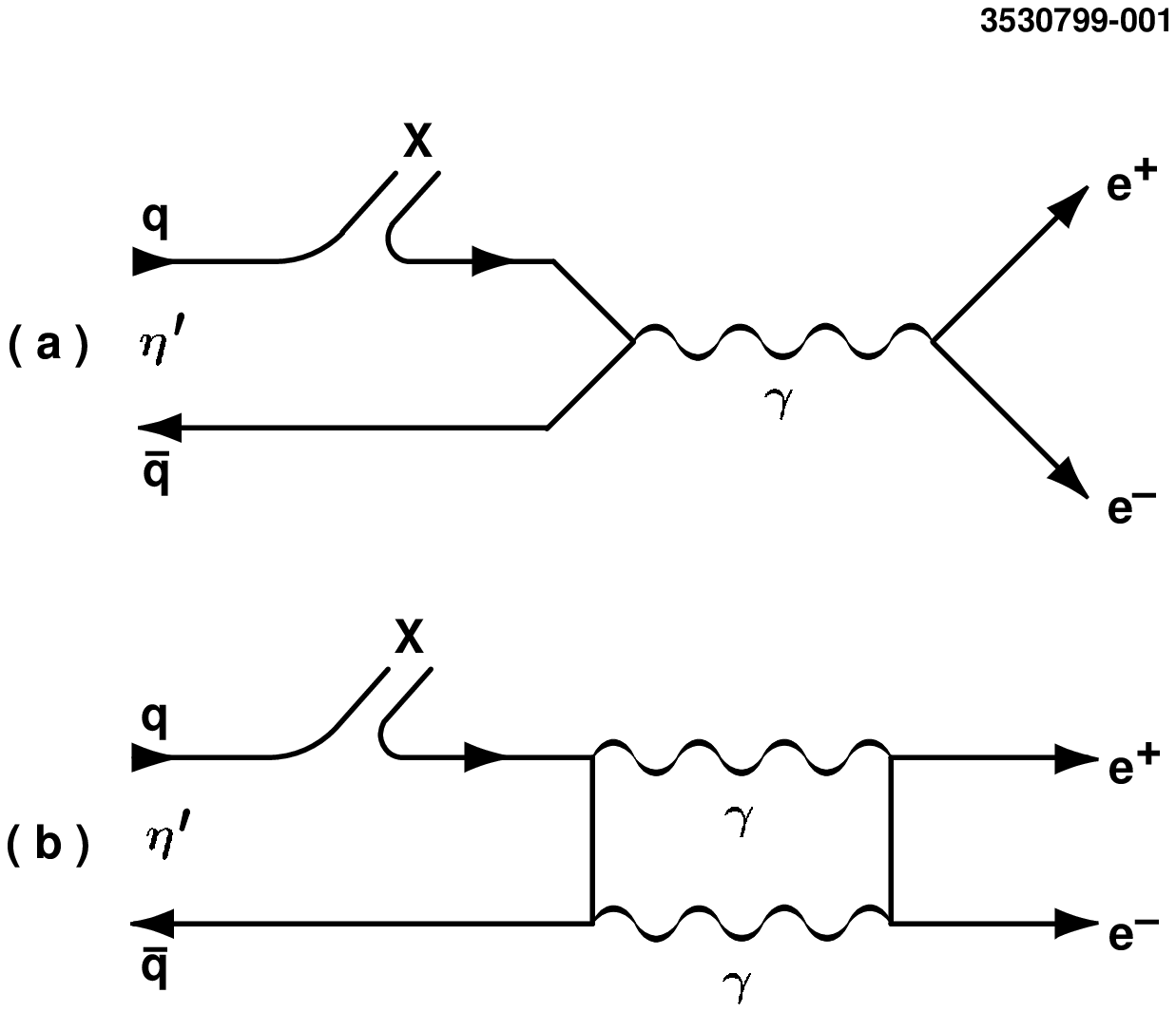, height=6.0in}}
\caption{(a) C-violating and (b) C-conserving contributions to the decay
$\eta ' \rightarrow e^{+}e^{-}X$, where $X$ is an $\eta$ or $\pi^{0}$.}
\label{feyn}
\end{figure}

The decays $\eta ' \rightarrow e^{+}e^{-}\eta$ and $\eta ' \rightarrow e^{+}
e^{-}\pi^{0}$ can occur with one or two intermediate virtual photons, as
shown in Figure \ref{feyn}.  The 
$\eta '$, $\eta$, and $\pi^{0}$ are even eigenstates of C, while a photon
is C-odd; thus the one-photon process will be C-violating and the two-photon
process C-conserving.  Because $B(\eta ' \rightarrow \pi^{0}\gamma\gamma)
< 8\times 10^{-4}$, and the rates for these decays should be orders of 
magnitude smaller \cite{cheng}, any signal at the 10$^{-4}$ level or larger
would signify a large C-violating contribution or other new physics.  The
current 90$\%$ confidence upper limits for these decays \cite{pdg} are 1.1$\%$
for $e^{+}e^{-}\eta$ and 1.3$\%$ for $e^{+}e^{-}\pi^{0}$.

The Dalitz decays of the $\eta$ and $\pi^{0}$ have both been measured, with
branching fractions two orders of magnitude smaller than the $\gamma\gamma$
decays of the two mesons.  Because $B(\eta ' \rightarrow \gamma\gamma)
= (2.12\pm 0.13)\%$ \cite{pdg}, we may expect the $\eta '$ Dalitz decay at
the 10$^{-4}$ level.  The decay $\eta ' \rightarrow e\mu$ with no 
accompanying neutrinos is an example of lepton number violation.  The 
theoretical upper bound for this decay is on the order of 10$^{-11}$, 
determined from experiments on $\mu^{-} \rightarrow e^{-}$ conversion
in heavy nuclei \cite{herczeg}.  A host of intermediate particles, including
massive neutrinos or leptoquarks \cite{white}, could manifest themselves
through a larger branching fraction for this decay.  No measured upper limit
has been published for either $\eta ' \rightarrow e^{+}e^{-}\gamma$ or
$\eta ' \rightarrow e\mu$. 

The common decay $\eta ' \rightarrow \pi^{+}\pi^{-}\eta$ $(\eta \rightarrow
\gamma \gamma)$ was used both as a normalization for our rare decay searches
and to investigate the structure of the $\eta '$.  Defining the Dalitz 
variables
\begin{equation}
y=[2+(m_{\eta}/m_{\pi})]{T_{\eta}\over Q} - 1,
\hspace{.15in}x={\sqrt{3}\over Q}(T_{\pi_{1}} - T_{\pi_{2}})
\end{equation}
in which $T$ represents kinetic energy in the $\eta '$ rest frame and 
$Q = m_{\eta '} - (m_{\eta} + 2m_{\pi})$, we fit the matrix element with
the Particle Data Group \cite{pdg} parameterization, $\mid$M$\mid ^{2} = 
A (\mid 1 + \alpha y \mid ^{2} + cx^{2})$.  Of particular interest is the real 
component of the complex constant
$\alpha$, which is a linear function of the kinetic energy of the $\eta$.  
A non-zero value of $\alpha$ may represent the contribution of a gluon 
component to the $\eta '$ decay \cite{alde}.  There are two published 
measurements of
Re$(\alpha)$; $-0.08\pm 0.03$ for $\eta ' \rightarrow \eta\pi^{+}\pi^{-}$ 
\cite{kalb} and $-0.058\pm 0.013$ for $\eta ' \rightarrow \eta\pi^{0}\pi^{0}$
\cite{alde}.  Though these measurements are consistent, $\alpha$ need
not be the same for the two decays.  

Data were collected using the CLEO II detector \cite{nim} at the Cornell
Electron Storage Ring (CESR).  We used 3.11 fb$^{-1}$ at the $\Upsilon(4S)$ 
resonance, 10.58 GeV, and 1.69 fb$^{-1}$ at 10.52 GeV.  Approximately 
$(15\pm 3)\%$ of our $\eta '$ sample comes from $B\bar{B}$ decays.  
Charged particle momenta are measured in a 67-layer tracking system immersed  
in a 1.5 T solenoidal magnetic field.  The main drift chamber also determines
a track's specific ionization ($dE/dx$), which aids in particle 
identification.  A 7800-crystal CsI calorimeter detects photons and is the
primary tool for electron identification.  Muons are identified using 
proportional counters placed at various depths in the steel return yoke of the 
magnet.

Selection criteria for $\eta ' \rightarrow \pi^{+}\pi^{-}\eta$ ($\eta 
\rightarrow \gamma \gamma$) were as follows.  We restricted our search to 
hadronic events by requiring at least 
five charged tracks in the drift chamber.  Photons from $\eta$ candidate decays
were required to
be in the region of best energy resolution in the calorimeter (80$\%$ of the
solid angle), to have an energy of at least 200 MeV, and not to overlap any
noisy crystals or the projections of charged tracks.  We rejected both photons
of pairs with an invariant mass within 12.5 MeV/$c^{2}$ of the $\pi^{0}$
mass.  Candidate pions must be well-tracked and must originate in the 
interaction region, within $\pm$3mm radially in the plane perpendicular to the 
beam and within $\pm$2cm in the beam direction.  Because we intend to use 
these decays
as normalization of our search for decays containing electrons, some
requirements are designed to reduce systematic uncertainties in electron
identification.  We veto tracks which have a vertex with another track in the
beam pipe or tracking chamber walls, because we will need to veto gamma 
conversions when seeking $\eta '$ decays involving electrons.  Similarly, 
tracks projecting to the calorimeter endcap are rejected, because electron 
identification degrades in this region.

\begin{figure}
\centerline{\epsfig{figure=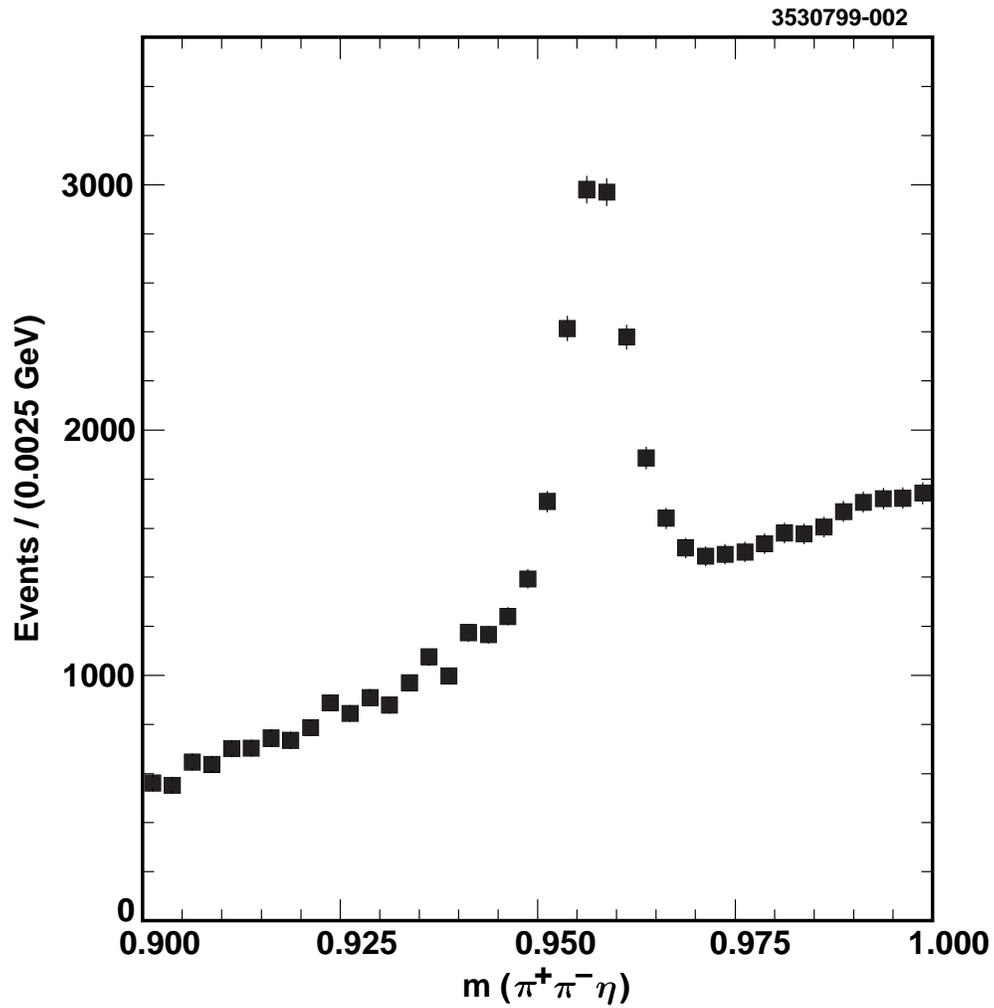, height=6.0in}}
\caption{Measured invariant mass distribution from CLEO II data for $\eta '
\rightarrow \pi^{+}\pi^{-}\eta$ ($\eta \rightarrow \gamma \gamma$).}
\label{norm}
\end{figure}

We kinematically fit each photon pair with invariant mass within 40 
MeV/$c^{2}$ of the $\eta$ mass.  By requiring
the $\eta$ and $\eta '$ to have minimum momenta of 0.6 and 1.0 GeV/$c$,
respectively, we limit random combinations of real $\eta$'s and 
charged pions, which are our largest background.  The final signal of roughly
6700 events is shown in Figure \ref{norm}.  From a Monte Carlo sample of 
$1.3 \times 10^{5}$ events we calculate an efficiency of $(2.96\pm 0.08)\%$,
indicating that about $1.3\times 10^{6}$ $\eta '$ mesons were produced in
our sample.

The searches for $\eta ' \rightarrow e^{+}e^{-}\eta$ ($\eta \rightarrow \gamma
\gamma$) and $\eta ' \rightarrow e^{+}e^{-}\pi^{0}$ ($\pi^{0} \rightarrow
\gamma \gamma$) use the same criteria as the normalizing mode, except that
we now require the charged tracks to be positively identified as electrons.
Our electron identification algorithm utilizes shower energy, track momentum, 
specific ionization loss,
and shower shape to achieve an efficiency of roughly $90\%$ with a fake rate 
from charged pions of less than 0.5$\%$.  Signal efficiency is estimated from
Monte Carlo samples for each decay, fitting each with a Gaussian and defining 
the signal region to be
$2\sigma$ to either side of the mean.  Average background is determined by
interpolating the data between 0.85 and 1.05 GeV/$c^{2}$ in invariant mass, 
excluding the signal region.

\begin{figure}
\centerline{\epsfig{figure=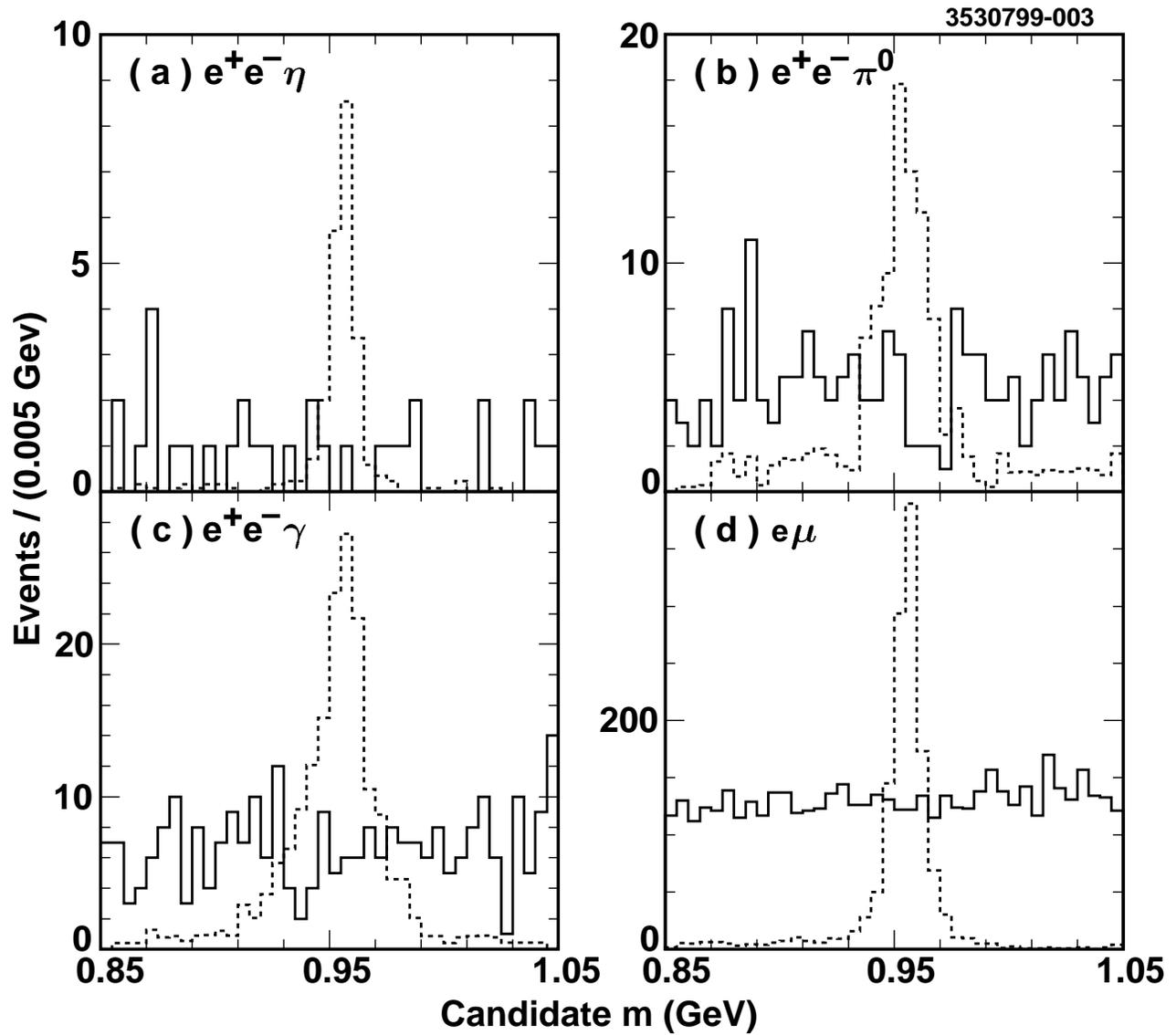, height=6.0in}}
\caption{Invariant mass distributions for $\eta '$ decays to (a) $e^{+}e^{-}
\eta$, (b) $e^{+}e^{-}\pi^{0}$, (c) $e^{+}e^{-}\gamma$, and (d) $e\mu$.  The
solid histogram represents the measured results, while the dashed curve is
the Monte Carlo prediction with arbitrary normalization.}
\label{data}
\end{figure}

The data (solid) and Monte Carlo (dashed) results for $e^{+}e^{-}\eta$ and 
$e^{+}e^{-}\pi^{0}$ are shown in Figures \ref{data}a  and \ref{data}b, 
respectively.  Neither shows any statistically significant signal in data.  
The efficiency for the 
$e^{+}e^{-}\eta$ mode is estimated from Monte Carlo to be $(0.27\pm 0.02)\%$.
There is one event in the signal region with an expected background of 3.0
events; Poisson statistics gives us a $90\%$ confidence upper limit of 2.8 
events.  The efficiency for the $e^{+}e^{-}\pi^{0}$ mode is $(0.37\pm 0.02)\%$.
There are 23 signal-region events and an average 
background of 31.0 events for an upper limit of 6.2 events at $90\%$
confidence level.  The background
for this mode is larger due to the greater number of neutral pions available
for random combinations.

The Monte Carlo samples used above distribute the decay products evenly over 
phase space.  One might argue that this is not appropriate for the one-photon
C-violating diagrams for these decays, and that we may be unknowingly cutting
away a signal of new physics.  To check this, we generated a second Monte
Carlo sample for each decay, using the measured electron-pair mass distribution
from the Dalitz decay of the $\eta$ \cite{jane}.  These Monte Carlo samples
give us estimated efficiencies of $(0.21\pm 0.02)\%$ for $e^{+}e^{-}\eta$ and
$(0.31\pm 0.03)\%$ for $e^{+}e^{-}\pi^{0}$.  For each mode we will use the
mean of the two Monte Carlo estimates as the efficiency and introduce half the
difference as a systematic uncertainty.

For the $\eta ' \rightarrow e^{+}e^{-}\gamma$ search we require a minimum 
photon energy of 0.6 GeV to
reduce combinatoric background.  We also see a significant background from
random photons combined with an electron pair from the Dalitz decays of a
$\pi^{0}$ or $\eta$.  We reduce this background by vetoing any event with an
$e^{+}e^{-}\gamma$ combination with invariant mass within $3\sigma$ (21
MeV/$c^{2}$) of the $\pi^{0}$ mass or within $2\sigma$ (26 MeV/$c^{2}$) of
the $\eta$ mass.  The data (solid) and Monte Carlo results are shown in 
Figure \ref{data}c; the Monte Carlo sample again uses the 
electron-pair mass distribution from
the $\eta$ Dalitz decay.  The efficiency for this mode is estimated to be
$(1.01\pm 0.06)\%$.  There is again no significant signal, with 51 
signal-region events and a background of 53.6 events, for a $90\%$ confidence 
upper limit of 11.7 events.

The criteria for muon identification in $\eta ' \rightarrow e\mu$ are 
identical to those for pions, with the additional requirement that the
muon penetrate three interaction lengths of material outside the calorimeter
(roughly a 1.0 GeV/$c$ momentum requirement).  The results of this search are 
shown
in Figure \ref{data}d.  The efficiency is found to be $(4.92\pm 0.15)\%$.  
There are 650 events with an average background of 672 events, for an upper 
limit of 30.1 events at $90\%$ confidence.

Table \ref{error} shows all sources of systematic uncertainty in our
analyses which do not cancel in the normalization to the $\eta '
\rightarrow \pi^{+}\pi^{-}\eta$ mode.  We used Poisson statistics to 
generate a
confidence level distribution for the number of signal events in each mode.  
We then ran a simulation which assigned an error to each probability in the
distribution, with the errors corresponding to a Gaussian with a width
determined by the total uncertainty for that mode.  This 'smearing' of
the distribution increased the number of signal events at the $90\%$ 
confidence level to the amount shown in Table \ref{smear} for each mode.
Normalizing the signal limit and efficiency for each mode to that of
$\eta ' \rightarrow \pi^{+}\pi^{-}\eta$ allows us to calculate $90\%$
confidence upper limits on the rare decay branching fractions:
\begin{equation}
B(\eta ' \rightarrow e^{+}e^{-}\eta) < 2.4\times 10^{-3},
\end{equation}
\begin{equation}
B(\eta ' \rightarrow e^{+}e^{-}\pi^{0}) < 1.4\times 10^{-3},
\end{equation}
\begin{equation}
B(\eta ' \rightarrow e^{+}e^{-}\gamma) < 0.9\times 10^{-3},
\end{equation}
\begin{equation}
B(\eta ' \rightarrow e\mu) < 4.7\times 10^{-4}.
\end{equation}

\begin{figure}
\centerline{\epsfig{figure=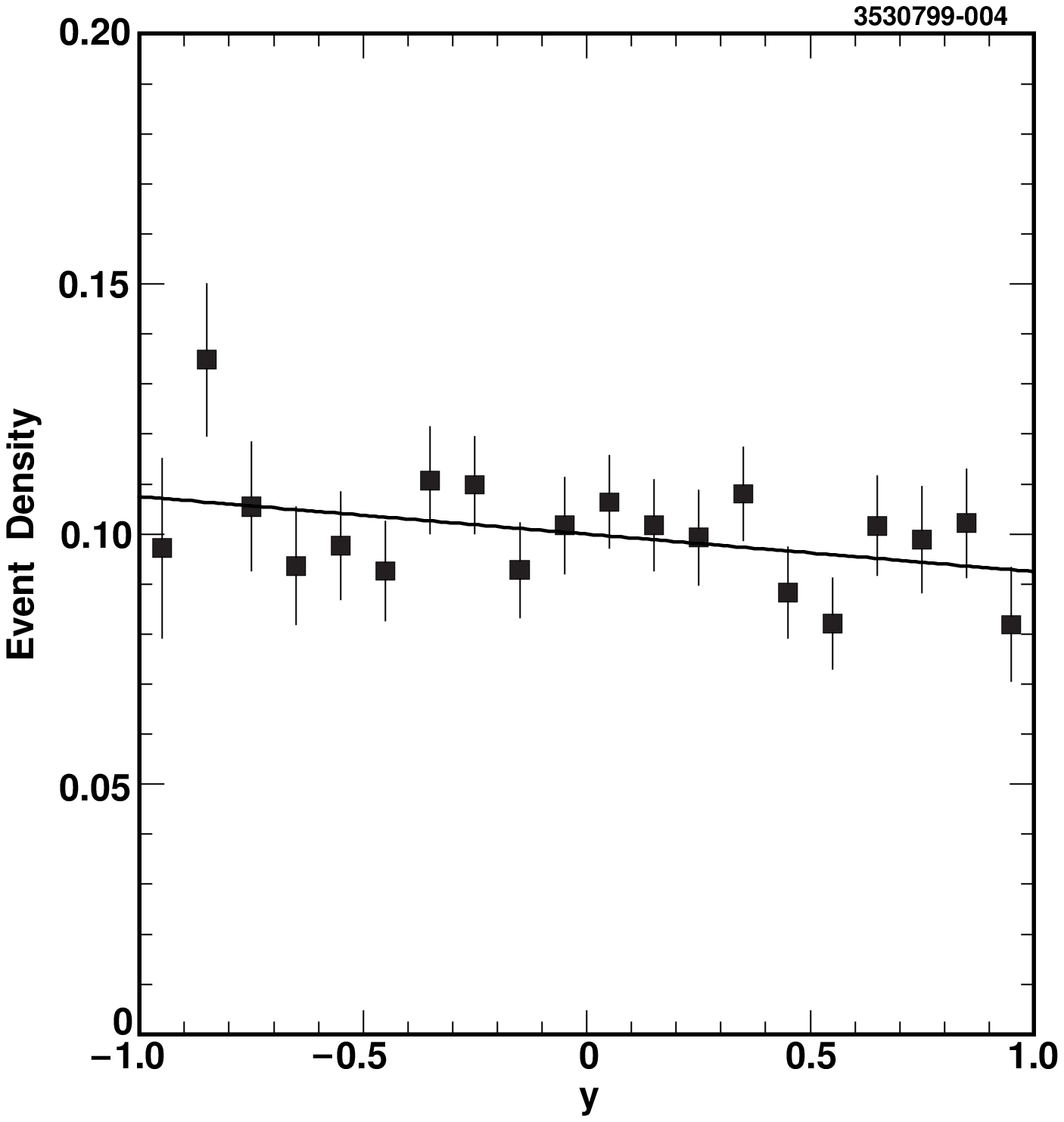, height=6.0in}}
\caption{Linear fit to the event density of the $\eta ' \rightarrow \eta 
\pi^{+}\pi^{-}$ Dalitz plot, projected along the y-axis, with $p_{\eta} >
0.6$ GeV/$c$ and sidebands defined as 4-6$\sigma$ to either side of the
mean.}
\label{alpha}
\end{figure}

In measuring Re$(\alpha)$ we use the selection criteria for $\eta ' 
\rightarrow \pi^{+}\pi^{-}\eta$ ($\eta \rightarrow \gamma \gamma$) with the
gamma conversion and calorimeter endcap requirements removed for greater
efficiency.  We fit the 
$\eta\pi^{+}\pi^{-}$ signal in data with a Gaussian and define the
signal region as $2\sigma$ to either side of the mean.  Background is removed
by sideband subtraction; we use
three different definitions of sideband ($3-5\sigma$, $4-6\sigma$, and 
$5-7\sigma$ from the mean).  Our sensitivity to the coefficient of $x^{2}$ is
small, so we plot the projection along the $y$-axis for each choice of
sideband.   Because the 
requirement on the $\eta$ momentum may be biasing the distribution of 
$T_{\eta}$, we also measure Re$(\alpha)$ with minimum $\eta$ momenta of 
0.4 GeV/$c$ and 0.8 GeV/$c$.  Efficiency corrections are made by dividing
our results by those from a Monte Carlo sample of $9.3\times 10^{5}$ $\eta '
\rightarrow \pi^{+}\pi^{-}\eta$ $(\eta \rightarrow \gamma \gamma)$ decays.
We fit the results
with a line whose slope will be twice the real component of $\alpha$.  One 
such plot is shown in Figure \ref{alpha}.  The slope varies by $\pm 0.012$ 
with choice of sideband, $\pm 0.006$ with choice of $\eta$ momentum 
requirement, $\pm 0.010$ from the statistics of the Monte Carlo sample, and 
$\pm 0.018$ due to the statistical uncertainty of the fit. 
Adding these uncertainties in quadrature yields a result of
Re $(\alpha) = -0.021\pm 0.025$.  While consistent with previous measurements,
this result is much smaller and is also consistent with zero.

In summary, 
we have searched for rare decays of the $\eta '$ involving at least one
electron in $e^{+}e^{-}$ collisions at the CLEO II detector.  We find
no statistically significant signals, and have assigned $90\%$ confidence
upper limits of $B(\eta ' \rightarrow e^{+}e^{-}\eta) < 2.4\times 
10^{-3}$, $B(\eta ' \rightarrow e^{+}e^{-}\pi^{0}) < 1.4\times 
10^{-3}$, $B(\eta ' \rightarrow e^{+}e^{-}\gamma) < 0.9\times 10^{-3}$, and
$B(\eta ' \rightarrow e\mu) < 4.7\times 10^{-4}$.  These measurements 
represent an order of magnitude improvement over previous limits, but are at 
least another order of magnitude short of discerning the Dalitz decay or 
suggesting a difficulty for the Standard Model.  We have also examined the
structure of the $\eta ' \rightarrow \pi^{+}\pi^{-}\eta$ and measured 
Re$(\alpha) = -0.021\pm 0.025$, a value consistent with but much smaller
than previous measurements.

We gratefully acknowledge the effort of the CESR staff in providing us with
excellent luminosity and running conditions.
This work was supported by 
the National Science Foundation,
the U.S. Department of Energy,
the Research Corporation,
the Natural Sciences and Engineering Research Council of Canada, 
the A.P. Sloan Foundation, 
the Swiss National Science Foundation, 
and the Alexander von Humboldt Stiftung.  

\pagebreak


\begin{table}
\begin{center}
\begin{tabular}{l c c c c}
Source & $e^{+}e^{-}\eta$ & $e^{+}e^{-}\pi^{0}$ & $e^{+}e^{-}\gamma$ &
$e\mu$\\ \hline
Photon detection efficiency & --- & --- & 3.0$\%$ & 6.0$\%$ \\
Electron ID efficiency & 6.0$\%$ & 6.0$\%$ & 6.0$\%$ & 3.0$\%$ \\
Muon ID efficiency & --- & --- & --- & 0.5$\%$ \\
$B(\eta ' \rightarrow \pi^{+}\pi^{-}\eta)$ & 0.7$\%$ & 0.7$\%$ & 0.7$\%$
& 0.7$\%$ \\
$B(\eta \rightarrow \gamma \gamma)$ & --- & 3.4$\%$ & 3.4$\%$ & 3.4$\%$ \\
$N_{\eta ' \rightarrow \pi^{+}\pi^{-}\eta}$ (fit to data) & 3.6$\%$ & 3.6$\%$ 
& 3.6$\%$ & 3.6$\%$ \\
$\varepsilon_{\eta ' \rightarrow \pi^{+}\pi^{-}\eta}$ (MC stat) & 2.5$\%$
& 2.5$\%$ & 2.5$\%$ & 2.5$\%$ \\
$\varepsilon$(rare decay) (MC stat) & 8.3$\%$ & 5.9$\%$ & 5.5$\%$ & 3.1$\%$ \\
Choice of MC model & 12.5$\%$ & 8.8$\%$ & --- & --- \\
$B\bar{B}$ contribution & 1.6$\%$ & 1.1$\%$ & 0.6$\%$ & 0.7$\%$ \\
\hline
Total & 16.8$\%$ & 13.4$\%$ & 10.3$\%$ & 9.3$\%$ \\
\end{tabular}
\end{center}
\caption{Summary of systematic uncertainties which do not cancel in
normalization.}
\label{error}
\end{table}

\begin{table}
\begin{center}
\begin{tabular}{c c c}
Rare & Limit without & Limit with\\ 
decay & systematics (events) & systematics (events)\\ \hline
$\eta ' \rightarrow e^{+}e^{-}\eta$ & 2.8 & 3.0\\
$\eta ' \rightarrow e^{+}e^{-}\pi^{0}$ & 6.2 & 6.4\\
$\eta ' \rightarrow e^{+}e^{-}\gamma$ & 11.7 & 11.9\\
$\eta ' \rightarrow e\mu$ & 30.1 & 30.4\\
\end{tabular}
\end{center}
\caption{Results of applying systematic uncertainty to confidence level
distribution.}
\label{smear}
\end{table}

\begin{references}
\bibitem{nefkens} B.M.K.~Nefkens, Few Body Syst. Suppl. 9, 193-202 (1995).
\bibitem{pdg} Particle Data Group, C.~Caso {\it {et al}}., Eur. Phys. 
J. C {\bf {3}}, 1 (1998).
\bibitem{cheng} T.P.~Cheng, Phys. Rev. {\bf {162}}, 1734 (1967).
\bibitem{herczeg} P.~Herczeg, in {\it {Rare Decays of Light Mesons}}, edited
by B.~Mayer (Editions Frontieres, Gif-sur-Yvette, 1990).
\bibitem{white} D.B.~White, Phys. Rev. D {\bf {53}}, 6658 (1996).
\bibitem{alde} D.~Alde {\it {et al}}., Phys. Lett. B {\bf {177}}, 115 (1986).
\bibitem{kalb} G.R.~Kalbfleisch, Phys. Rev. D {\bf {10}}, 916 (1974). 
\bibitem{nim} Y.~Kubota {\it {et al}}., Nucl. Instrum. Methods Phys. Res.,
Sect. A {\bf {320}}, 66 (1992).
\bibitem{jane} M.R.~Jane {\it {et al}}., Phys. Lett. B {\bf {59}}, 103 (1975).
\end{references}
\end{document}